# Circular-ribbon flares and the related activities

## Qingmin Zhang[1,2*]

[1*]Key Laboratory of Dark Matter and Space Astronomy, Purple Mountain Observatory, Chinese Academy of Sciences, Nanjing, 210023, China.
[2]Yunnan Key Laboratory of Solar Physics and Space Science, Kunming, 650216, China.

Corresponding author(s). E-mail(s): zhangqm@pmo.ac.cn;

## Abstract

Solar flares are an impulsive increase of emissions as a result of impetuous release of magnetic free energy. In this paper, I will present the recent progress on circular-ribbon flares (CRFs) and their related activities, including coronal jets, filaments, coronal mass ejections (CMEs), radio bursts, coronal dimmings, and coronal loop oscillations. Owing to the prevalence of three-dimensional (3D) magnetic null points and the corresponding fan-spine topology in the solar atmosphere, CRFs are regularly observed in ultraviolet (UV), extreme-ultraviolet (EUV), and Hα passbands. Spine reconnection and fan reconnection around the null points are predominantly responsible for the energy release and subsequent particle acceleration. Slipping reconnection at quasi-separatrix layers (QSLs) may explain the sequential brightening or rapid degradation of the circular ribbons. Periodic or quasi-periodic acceleration and precipitation of nonthermal particles in the chromo- sphere produce observed quasi-periodic pulsations (QPPs) of CRFs in multiple altitudes as well as wavelengths. Like two-ribbon flares, the injected high-energy particles result in explosive evaporation in circular and inner ribbons, which is characterized by simultaneous blueshifts in the coronal emission lines and redshifts in the chromospheric emission lines. Homologous CRFs residing in the same active region present similar morphology, evolution, and energy partition. The peculiar topology of CRFs with closed outer spines facilitates remote brightenings and EUV late phases, which are uncommon in two-ribbon flares. Besides, CRFs are often accompanied by coronal jets, type III radio bursts, CMEs, shock waves, coronal dimmings, and kink oscillations in coronal loops and filaments. Magnetohydrodynamics numerical simulations are very helpful to understand the key problems that are still unclear up to now. Multiwavelength and multipoint observations with state-of-the-art instruments are enormously desired to make a breakthrough. The findings in CRFs are important for a comprehensive understanding of solar flares and have implication for stellar flares.

**Keywords:** Solar flares, Solar magnetic fields, Magnetic reconnection, Particle acceleration

# 1 Introduction

Solar flares are impulsive increases of emissions observed in multiple wavelengths (Carrington, 1859; Hudson, 1991; Gosling, 1993; Fletcher et al, 2011; Kontar et al, 2011; Shibata and Magara, 2011; Janvier et al, 2015; Benz, 2017). The prestored magnetic free energy in active regions (ARs) is impetuously released and converted into radiation, thermal and kinetic energies of particles within a few minutes to a few hours (Priest and Forbes, 2002; Yu et al, 2020). It is generally believed that the release and conversion of free energy is by means of magnetic reconnection, during which magnetic field lines with opposite polarities come into contact, break, and reconnect (Chen et al, 1999; Yokoyama and Shibata, 2001; Shiota et al, 2005; McLaughlin et al, 2009; Yamada et al, 2010). In the framework of standard two-dimensional (2D) flare model, i.e., CSHKP model (Carmichael, 1964; Sturrock, 1966; Hirayama, 1974; Kopp and Pneuman, 1976), as a filament or flux rope erupts upward to drive a coronal mass ejection (CME; Forbes, 2000; Chen, 2011), a thin, long current sheet is spontaneously formed underneath, which is usually observed in extreme-ultraviolet (EUV) (Savage et al, 2012a; Cheng et al, 2018; Chitta et al, 2021) and white light (Ko et al, 2003; Lin et al, 2005; Liu, 2013; Bem- porad et al, 2022). With the development of high-resolution and high-cadence observational facilities, such as the New Vacuum Solar Telescope (NVST; Liu et al, 2014) and the Atmospheric Imaging Assembly (AIA; Lemen et al, 2012) on board the Solar Dynamics Observatory (SDO; Pesnell et al, 2012) spacecraft, direct imaging observations of inflow and outflow in current sheets are substantial (Yokoyama et al, 2001; Takasao et al, 2012; Su et al, 2013; Yang et al, 2015b; Li et al, 2016; Xue et al, 2016; Yan et al, 2018, 2022). Spectroscopic observations reveal large nonthermal velocity in current sheets (Hara et al, 2011; Li et al, 2018b). The fast outflow reaches and collides with the post- flare loops, generating the so-called supra-arcade downflows (SADs; McKenzie, 2000; Savage et al, 2012b; Liu, 2013; Wang et al, 2023b) observed in EUV and soft X-ray (SXR) wavelengths.

The nonthermal electrons accelerated by magnetic reconnection stream downward to the dense chromosphere and lose their energy to generate localized heating. Meanwhile, HXR emissions are produced at the footpoints of flare loops as a result of bremsstrahlung (Krucker et al, 2008; White et al, 2011; Krucker and Battaglia, 2014). Apart from the conjugate footpoint sources, a loop-top source and above-the-loop-top source are occasionally detected (Masuda et al, 1994; Sui and Holman, 2003) and investigated (Kong et al, 2022). Two bright ribbons, mapping the footpoints of flare loops, show up in white light, Hα, Ca II 396.8 nm H line, UV, EUV, and near-infrared (e.g., He I 10830 Å) (Krucker et al, 2011a). The exceedingly hot and dense materials expand upward and create strong emissions in SXR and EUV passbands, known as chromospheric evaporation (see Section 4). As time goes by, the flare loops cool down due to radiative and conductive losses. The pre-flare phase, impulsive phase, and decay phase of flares are generally synchronized with the initial phase, acceleration phase, and propagation phase of the associated CMEs (Zhang et al, 2001, 2004). Both flares and CMEs are likely to have significant geoeffectiveness (Gopalswamy et al, 2007; Temmer, 2021).

Circular-ribbon flares (CRFs), as its name implies, are a special type of flares whose outer ribbons have closed circular or elliptic shapes. For the first time, Masson et al (2009) studied a C8.6 CRF in AR 10191 observed by the Transition Region And Coronal Explorer (TRACE; Handy et al, 1999) on 2002 November 16. The top panels of Figure 1 show two snapshots of the flare observed in UV 1600 Å. In Figure 1(a2), the flare ribbons consist of a shorter, inner ribbon (RA), a longer, circular ribbon (RC), and a straight remote ribbon (RB). In Figure 1(a1), the UV intensities of flare ribbons reach their maxima around 13:58:07 UT. Before the TRACE mission, ground-based solar telescopes had observed CRFs. Using the full-disk Hα and He I D3 images recorded by the Big Bear Solar Observatory (BBSO) in 1991, Wang and Liu (2012) noticed five CRFs in association with homologous jets (surges). Circular filaments are identified before eruption and the generation of flares. Liu et al (2013a) explored the M6.3 CRF on 1984 May 22 and estimated the radiated energy in D3 during the main peak, which is ∼$10^{30}$ erg. Using the Hα flare spectra obtained with the Domeless Solar Telescope (DST) at Hida Observatory in Japan, Ichimoto and Kurokawa (1984) analyzed the strong Hα red asymmetry, which is an indication of downward motion at speeds of 40∼100 km s⁻¹ in the flare chromosphere. The flare ribbons on 1982 June 20 and June 21 are composed of a circular ribbon and a compact inner ribbon (see their Fig. 1).

In contrast to the long history and plentiful observations of two-ribbon flares, investigations of CRFs are not so copious. However, remarkable progress has been made in this area since a growing number of sophisticated solar telescopes have come into play. In this brief review, I will describe the basic properties of CRFs in Section 2, particle acceleration in Section 3, energy transport and partition in Section 4, and their related activities in Section 5, including remote brightenings, EUV late phases, coronal jets and type III radio bursts, CMEs and shock waves, coronal dimmings, and kink oscillations in coronal loops and filaments. Discussions and open questions are presented in Section 6. Finally, a brief summary is given in Section 7.

# 2 Basic properties of CRFs

## 2.1 Magnetic topology and  reconnection

Magnetic null points (**B**=0) are ubiquitous not only in solar atmosphere (Demoulin et al, 1994; Brown and Priest, 2001; Longcope, 2005; Liu et al, 2011; Mandrini et al, 2014; Edwards and Parnell, 2015; Chen et al, 2016;  Titov et al, 2017; Wyper et al, 2021), but also in Earth's magnetosphere (Xiao et al, 2006; Fu  et al, 2015). Most of the null points occur at low latitudes of   the quiet Sun and the column of null points per square megameter above a height of 1.5 Mm is  3.1+/-0.3  Mm$^{-2}$  (Longcope and Parnell, 2009). Using a three-dimensional (3D) zero-$\beta$ magnetohydrodynamics (MHD) simulation, Török et al (2009) uncovered the formation of a 3D null point and the related fan-spine configuration. The null point is created at the interface of the emerging and pre-existing fields (Shibata et al, 1994; Moreno-Insertis et al, 2008). The formation of fan-spine structure needs two-step magnetic reconnections. The first reconnection occurs between emerging flux tubes and the outer coronal arcade, which generates a torsional MHD wave and a sheared loop system. The second reconnection occurs between the newly formed loops and the inner coronal arcade, which generates a second loop system. The topology near a 3D null point is determined  by  the three eigenvalues ($\lambda_1$, $\lambda_2$, and $\lambda_3$) of the Jacobian matrix (**J**), where     $J_{ij} = \partial B_i / \partial x_j$ (Lau and Finn, 1990; Parnell et al, 1996). For three real eigen- values, they are classified into two types: type A ($\lambda_1 > 0$,  $\lambda_2$  < 0,  $\lambda_3$  < 0)  and type B ($\lambda_1 < 0$, $\lambda_2 > 0$, $\lambda_3 > 0$). For one real ($\lambda_1$) and two  complex ($\lambda_2$,  $\lambda_3$) eigenvalues, the two types become A$_S$ ($\lambda_1 > 0$) and B$_S$ ($\lambda_1 < 0$), where S means spiral.

The 3D magnetic topology of CRFs is characterized by a null point and the associated dome-like fan-spine structure (Masson et al, 2009). Magnetic field lines approach (recede from) the null point along the spine and recede from (approach) the null point along the fan surface. The whole spine consists of two parts: an outer spine and an inner spine (see Figure 1). The inner spine under the null point is shorter in length. The outer, longer spine could be closed, connecting to a remote site (Yang et al, 2015a; Xu et al, 2017; Romano et al, 2017; Chen et al, 2019; Zhong et al, 2019, 2021; Joshi et al, 2021; Ning et al, 2022; Zhang et al, 2021b; Duan et al, 2022). Otherwise, the outer spine could be open, extending far into the interplanetary space (Pariat et al, 2009, 2010; Sun et al, 2012; Wang and Liu, 2012; Zhang et al, 2012; Joshi et al, 2018; Wyper et al, 2018; Guo et al, 2019). In consequence, the local space is distinctly divided into two domains with different magnetic connectivity: one inside the fan surface and the other outside the fan surface. Aside from the fan and spine, quasi-separatrix layer (QSL; Demoulin et al, 1996; Titov et al,

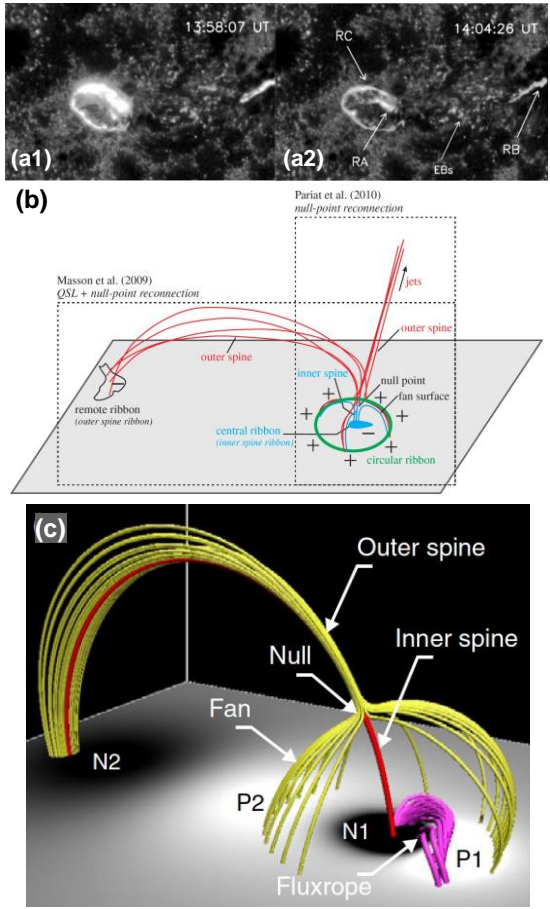

**Fig. 1** (a1-a2) A C8.6 CRF observed by the TRACE spacecraft in UV 1600 Å on 2002 November 16 (credit: Masson et al (2009)). (b) A combination of the schematic illustrations of the relationship among various activities (credit: Wang and Liu (2012)). (c) A model for flux rope eruption in the fan-spine topology (credit: Sun et al (2013)).

2002) field lines are found to be related to the fan, inner spine, and out spine (Masson et al, 2009; Reid et al, 2012; Yang et al, 2015a; Pontin et al, 2016; Hou et al, 2019). QSLs represent locations where magnetic connectivity changes abruptly, electric current is built up, and slip-running reconnection preferably occurs (Aulanier et al, 2007; Dud́ık et al, 2014; Li and Zhang, 2015).

In the photosphere, the magnetic polarity where the inner spine is rooted in is surrounded by the opposite polarity (Antiochos, 1998; Wang and Liu, 2012; Zhang et al, 2016b). Along the polarity inversion line (PIL), curved filaments or magnetic flux ropes may exist beneath the dome before eruption (Jiang et al, 2013a; Sun et al, 2013; Yang et al, 2015a; Xu et al, 2017; Song et al, 2018; Yang et al, 2020a,b; Zhang and Zheng, 2020). The filament rises and squeezes the null point to form a thin current sheet where breakout-type magnetic reconnection and particle acceleration take place (Pontin et al, 2013; Wyper et al, 2018). Nonthermal electrons flow down the reconnected field lines into the chromosphere and heat the localized plasma to generate the bright circular ribbon and inner ribbon in Ca II H, Hα, UV, and EUV wavelengths. Interestingly, dark circular ribbon is detected in the He I D3 line during the M6.3 CRF on 1984 May 22 (Liu et al, 2013a). The dark ribbon mirrors the bright ribbons simultaneously observed in Hα and He I 10830 Å. The filament eruption would be confined when the outer spine is connected back to the solar surface (Wyper and DeVore, 2016; Zhang et al, 2016b; Hernandez-Perez et al, 2017; Li et al, 2018a; Song et al, 2018; Zhang et al, 2021b; Mitra et al, 2022). Owing to the asymmetry of a fan-spine topology, the eruption tends to be nonradial (Jiang et al, 2013a). Sometimes, the eruption is successful to create a CME (Liu et al, 2015a; Joshi et al, 2015, 2017; Jiang et al, 2016). Apart from the inner and circular ribbons, a short pair of flare ribbons or kernals are detected beneath the ascending filaments (Joshi et al, 2015; Zhang et al, 2016a; Hernandez-Perez et al, 2017; Xu et al, 2017; Devi et al, 2020). To figure out the cause of escape of flare-accelerated particles, Masson et al (2013) carried out 2.5D numerical simulations of large-scale flux rope eruptions generating CMEs. A fan-spine structure is located near the open field of a coronal hole. A flux rope beneath the null point rises up and produces a standard two-ribbon flare, which accelerate particles. Once the flux rope touches the null point during eruption, interchange magnetic reconnection takes place between the flux rope and open field. Afterwards, the nonthermal particles previously trapped in the flux rope escape into the solar wind along the open field. The occurrence rate of filament or flux rope eruptions with CRFs is 38%, and the association rates increase with flare classes (Zhang et al, 2022d).

Theoretical research on magnetic reconnection in a null point is abundant. Particle-in-cell (PIC) experiment reveals that 3D null points are preferential sites of energy release, which is more efficient than in traditional 2D current sheets (Olshevsky et al, 2013). Nayak et al (2019) carried out data-constrained MHD simulation of a blowout jet and a confined C1.2 flare on 2016 December 5. The complex magnetic topology of the events consists of a pair of null points, a single null point, and a QSL. It is found that the generation of jet and flare ribbons are due to magnetic reconnections at the null point and a QSL after the eruption of the underlying mini-filament. Kumar et al (2021b) performed 3D MHD simulations to investigate magnetic reconnections in the presence of two magnetic nulls and QSLs. In the first case of a lower dome, strong electric currents and torsional fan reconnection are identified at the nulls, while weak QSL reconnection occurs at the hyperbolic flux tube. In the second case of a larger dome, QSL reconnection becomes stronger than that in the first case.

Kinetic reconnections via spine-aligned current and fan-aligned current are explored by Pontin et al (2004) and Pontin et al (2005), respectively. The formation of singular current layers at nulls in a line-tied volume during an ideal relaxation has been demonstrated numerically (Pontin and Craig, 2005). Pontin and Galsgaard (2007) investigated the current build-up and magnetic reconnection at a 3D null point, which is disturbed by rotations and shears. The location and direction of the current sheet are quite different under the two circumstances. Moreover, Pontin et al (2011) compared the locations and reconnection rates between torsional spine reconnection and torsional fan reconnection, uncovering the dependence of peak current and peak reconnection rate on the degree of asymmetry. For a detailed description of magnetic reconnection at 3D null points, see these review papers (Priest and Titov, 1996; Priest and Pontin, 2009; Pontin, 2011; Pontin and Priest, 2022; Lee, 2022).

## 2.2 Location, area, radius, lifetime, and peak SXR flux

The sizes of circular ribbons have a wide range, from small sizes of 18 22 Mm (Wang and Liu, 2012; Xu et al, 2017; Li and Yang, 2019) to a middle size of 61 Mm (Ning et al, 2022) and a large size of 145 Mm, which is comparable to the diameter of Jupiter (Chen et al, 2019). An extremely huge size of $\sim$283 Mm was reported by Joshi et al (2017). Zhang et al (2022d) carried out a comprehensive statistical analysis of 134 CRFs from 2011 September to 2017 June, including 4 B-class, 82 C-class, 40 M-class, and 8 X-class flares. Among the 134 events, 97 and 37 are confined and eruptive events, respectively. In Figure 2, panel (a) shows the locations of CRFs, all of which are in ARs. Panel (b) shows the distribution of latitude, which is accordant with that of normal flares (Christe et al, 2008; Ajello et al, 2021). Panel (c) shows the distribution of flare area ($A_{CF}$) after correcting the projection effect, which is perfectly fitted with a log-normal function (red curve):

$$\frac{1}{N}\frac{dN}{dx} = f(x, \mu, \sigma) = \frac{1}{x\sigma\sqrt{2\pi}}e^{-\frac{(\ln x - \mu)^2}{2\sigma^2}}, x > 0, \qquad (1)$$

where $\mu = 6.51$ and $\sigma = 0.66$. Panel (d) shows the distribution of corresponding radius of the circular ribbons assuming a semisphere, $r_{CF} = \frac{A_{CF}}{\pi}$. Panel (e) shows the distribution of flare lifetime ($\tau_{CF}$), which could also be fitted with a log-normal function, where $\mu = 3.60$ and $\sigma = 0.64$. The median values of $A_{CF}$, $r_{CF}$, and $\tau_{CF}$ are $\sim$631 Mm², $\sim$14 Mm, and $\sim$35 minutes, respectively. A positive correlation is found between $A_{CF}$ and $\tau_{CF}$, suggesting that larger CRFs have longer lifetimes. This could be explained by the linear correlation between the cooling timescale by conductive loss and flare area $A_{CF} = c_0 L^2$, where $c_0$ is a constant. Panel (f) shows the distribution of peak SXR flux in 1 8 Å, which could be fitted with a power-law function with a power index of $a = -1.42$. Power-law distribution is prevalent not only in solar flares (Crosby et al, 1993; Aschwanden et al, 2000; Christe et al, 2008; Lu et al, 2021a), but also in stellar flares (Maehara et al, 2012; Shibata et al, 2013).

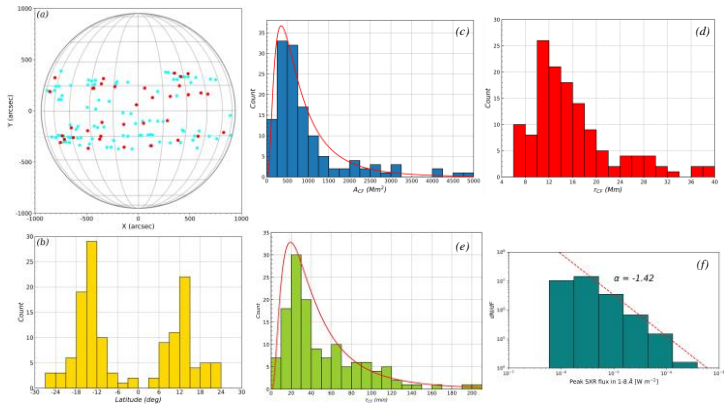

**Fig. 2** Statistical properties of CRFs, including the locations (a), distributions of latitude (b), area (c), equivaient radius (d), lifetime (e), and peak SXR flux in 1   8 Å (credit: Zhang et al (2022d)).

## 2.3 Ribbon motion

For two-ribbon flares, diverse motions have been found in the ribbons, such as contraction, separation, and elongation observed in UV, EUV, and H$\alpha$ wavelengths (Qiu et al, 2002; Ji et al, 2004, 2006; Li and Zhang, 2015; Xu et al, 2016; Qiu et al, 2017). The speed of separation and line-of-sight (LOS) magnetic field strength can be used to derive the magnetic reconnection rate of a flare (Qiu et al, 2002; Cannon et al, 2023). Recently, it is uncovered that fast sweeping motion of a flare ribbon is an indication of complete replacement of the footpoints of an erupting flux rope, whose toroidal flux increases from $3.3 \times 10^{20}$ Mx to $7.9 \times 10^{20}$ Mx (Gou et al, 2023).

For CRFs, the brightenings of ribbons are not always in phase. Instead, the outer ribbons of CRFs show sequential brighenings in clockwise (Shen et al, 2019b) or counterclockwise (Masson et al, 2009; Liu et al, 2013a; Li et al, 2017d; Xu et al, 2017) directions. Li et al (2018a) studied an M1.8 confined CRF, which experienced two episodes of magnetic reconnections as a result of multiple filament eruptions on 2016 February 13. During both episodes of eruption, the circular ribbon shows elongation motion in the counterclockwise direction. The apparent speeds of elongation in the chromosphere are $210-220$ km s$^{-1}$ and $60-70$ km s$^{-1}$, respectively. It is concluded that continuous null-point reconnections give rise to sequential opening of the filament and rapid shifts of the fan plane footpoint. Shen et al (2019b) reported an interesting round-trip slipping motion of the western part of outer circular ribbon within $\sim$2 minutes on 2014 July 31. The ribbon first undergoes northward motion and then experiences southward motion. Zhang et al (2020c) reported fast degradation of the outer circular ribbon in the counterclockwise direction during the impulsive and decay phases of the C5.5 CRF on 2014 August 24. The degradation motions are in phase in all UV and EUV wavelengths of SDO/AIA. The  speeds of degradation decrease from $58-69$ km s$^{-1}$ to $29-35$ km s$^{-1}$, suggesting a deceleration. Meanwhile, the flare-related jet undergoes untwisting motion in the same direction. It is concluded that the coherent brightenings and subsequent slipping motion of the circular ribbon result from impulsive null-point reconnection and continuous slipping reconnection, respectively. Pontin et al (2013) explored spine-fan reconnection at a 3D null point and found flux trans- fer across the current sheet forming around the null point. The flipping of magnetic field lines may explain the observed fast degradation of the circular ribbon (Zhang et al, 2020c).

# 3 Particle acceleration in CRFs

Particle acceleration at a 3D reconnection site has been extensively investigated. In response to continuous shearing motions of the spine, a thin (but finite) current sheet is formed at the null point and parallel electric field grows during magnetic reconnection (Pontin et al, 2007). Using the test-particle method, Dalla and Browning (2005) derived the trajectories of protons around a 3D null point during spine reconnection. Strong acceleration of protons is realized in the strong electric field regime ($E$ = 1.5 kV m$^{-1}$). Meanwhile, the energy gain is quite sensitive to the location of injection into the simulation box. In a follow-up work, Dalla and Browning (2006) discovered that two energetic populations of particles are generated. The trapped particles remain near the null point, while the escaping particles leave the null point in two symmetric jets along field lines near the spine, the latter of which may explain the type III radio bursts associated with CRFs (Zhang et al, 2016a, 2021b). Apart from spine reconnection, there is fan reconnection at a 3D null point, which seems less efficient than spine reconnection in supplying seed particles to the diffusion region of strong electric field (Dalla and Browning, 2008). The maximum energy of protons reaches 1 MeV and 30 MeV in fan and spine reconnections, respectively. Moreover, no jets are produced and particles escape in the fan plane in fan reconnection, implying that fan reconnection is irrelevant in particle acceleration during CRFs. On the contrary, Stanier et al (2012) concluded that protons can be accelerated up to 0.1 GeV by the electro-magnetic fields of fan reconnection observed in flares, while spine reconnection is less efficient. Guo et al (2010) investigated the role of convective electric field in accelerating particles near a 3D null point. It is revealed that pro-tons could more easily be accelerated than electrons, suggesting that resistive electric field parallel to the magnetic field is needed to efficiently accelerate electrons. Besides, accelerations of electrons and protons in anemone- like coronal jets are exquisitely explored using the test-particle or PIC methods (Rosdahl and Galsgaard, 2010; Baumann and Nord- lund, 2012; Pallister et al, 2021). For an exhaustive demonstration of particle acceleration in solar flares, see these review papers (Aschwanden, 2002; Holman et al, 2011; Zharkova et al, 2011; Guo et al, 2020).

Quasi-periodic pulsations (QPPs) are widespread in astrophysics, such as fast radio bursts (FRBs; Niu et al, 2022; Xie et al, 2023), microquasars (Tian et al, 2023), solar and stellar flares (Van Doorsselaere et al, 2016; Zimovets et al, 2021). QPPs in solar flares are frequently detected in HXR (Kane et al, 1983; Zimovets and Struminsky, 2009; Kolotkov et al, 2018; Shi et al, 2023), SXR (Kumar et al, 2015), EUV (Li et al, 2015a), UV (Li and Zhang, 2015; Li et al, 2020d; Lu et al, 2021b), and radio (Yuan et al, 2019; Li et al, 2020c) wavelengths. The periods of QPPs have a wide range from tens of millisecond to several minutes (Kliem et al, 2000; Karlický et al, 2020; Dennis et al, 2017). Occasionally, multiple periods exist in a single event (Inglis and Nakariakov, 2009; Ning et al, 2022; Zhang et al, 2022b). QPPs may appear in the pre-flare phase (Li et al, 2020b), impulsive phase, and decay phase (Hayes et al, 2016) of flares. QPPs are clear indication of dynamic magnetic reconnection and periodic particle acceleration (Cheng et al, 2018; Clarke et al, 2021).

Using the SXR flux at 6–12 keV observed with the Gamma-ray Burst Monitor (GBM) on board the Fermi spacecraft, Kumar et al (2015) found QPP in a C-class CRF on 2013 July 20. The period and damping time are 202 s and 154 s, respectively. Meanwhile, propagating EUV disturbances are found to bounce back and forth at a speed of 560 km s$^{-1}$ between the footpoints of the hot arcade loops in 94 and 131 Å. The periods and damping time are 409 s and 1121 s, respectively. The propagating disturbances are explained by global standing slow magneto-acoustic wave excited by the flare (Fang et al, 2015; Mandal et al, 2016; Xia et al, 2022). Using multiwavelength observations on 2011 September 23, Kumar et al (2016b) analyzed an M1.9 class CRF, which shows QPP in HXR, radio, and EUV passbands. The period ( 3 min­utes) of QPP is close to the period of nearby sunspot oscillation as well as the period of filament rotation. It is concluded that the QPP results from repetitive reconnection and particle acceleration at the magnetic null point. Chen et al (2019) investigated an M8.7 CRF on 2014 December 17. Interestingly, QPPs with different periods appear at different locations. Pre-flare QPP with a period of 3 minutes is detected in the circular flare ribbon observed in AIA 171 and 304 Å. QPP with a period of 4 minutes is detected in the center of AR 12242 observed in AIA 1600 Å during the pre-flare and impulsive phase. Besides, QPP with a period of ∼2 minutes is exclusively detected in the flare site observed in 1-2 GHz during the impulsive phase. This event nicely demonstrates the complex behavior of QPPs in solar flares. Mitra et al (2022) studied an M4.0 flare as a result of failed flux rope eruption from below a magnetic null point on 2011 September 26. Before the onset of the flare, multiple episodes of brightenings ( 10 peaks) from the null point are detected, indicating intermittent null-point reconnection in the precursor phase, which may contribute to the gradual destabilization of the flux rope. Kashapova et al (2020) reanalyzed the M1.0 CRF on 2014 March 5. The ribbons include an inner ribbon (R1), a circular ribbon (R2), and a remote ribbon (R3). QPPs in H$\alpha$ emission with periods around 150 s, 125 s, and 190 s are detected. The period of 150 s is simultaneously observed in H$\alpha$, HXR, and microwave (4, 5.7, and 8 GHz), which is interpreted by periodic slipping reconnection in the fan surface. In the C3.1 CRF on 2015 October 16, Zhang et al (2016a) found QPP with periods of 32 42 s in UV and HXR wavelengths. It is conjectured that null-point reconnection and particle acceleration are modulated by a standing fast-mode wave in the flare loops. Thurgood et al (2017) performed 3D numerical simulation of oscillatory magnetic reconnection around a 3D null point. It is unveiled that oscillatory reconnection is self-generated after the collapse of the null point due to an external MHD wave, which is plausible to explain the observed QPPs in CRFs hosting null points. Using the Bifrost code, Nóbrega-Siverio and Moreno-Insertis (2022) carried out a 2D numerical experiment for coronal bright points (CBPs) where a coronal null point is included. Similarly, the reconnection at the null point is intermittent and oscillatory, with the angle of current sheet changes quasi-periodically between -45° and 45°. Li et al (2020b) studied the QPPs consisting of 7 well identified pulses in the pre-flare phase of an M1.1 CRF on 2015 October 16. The period is ∼ 9.3 minutes. The pulses, detected in Hα, SXR, and EUV (211 Å) wavelengths, are probably modulated by LRC-circuit oscillation in the current-carrying flare loop (Zaitsev et al, 1998). Ning et al (2022) analyzed three kinds of oscillation in the X1.1 CRF on 2013 November 10. The first kind is QPP with a period of ∼ 20 s detected in X-ray, EUV, and microwave passbands. The second type is periodic coronal jets related to the flare with a period of ∼ 72 s. The third type is coronal loop oscillation with a period of ∼ 168 s, which is probably induced by the flare-generated blast wave (Nakariakov et al, 1999). Figure 3 shows four examples of CRFs with QPPs.

Though imaging observations of two-ribbon flares in radio wavelengths are abundant (Gary et al, 2018; Fleishman et al, 2020), radio observations of CRFs are uncommon. Kumar et al (2016b) studied the M1.9 jet-related flare probabaly associated with a fan-spine topology in AR 11302 on 2011 September 23-24. The radio source at 17 GHz is cospatial with the flare region and the two HXR sources at 25 50 keV are cospatial with two flare kernels in 1600 Å. Likewise, Zhang et al (2019b) obtained the radio image at 17 GHz of the C5.5 CRF on 2014 August 24. The two bright patches (BP1 and BP2) are exactly consistent with the locations of inner and outer ribbons. Using microwave observations with the broadband Mingantu Spectral Radioheliograph (MUSER; Yan et al, 2009) as well as the Nobeyama Radioheliograph (NoRH; Nakajima et al, 1994), Chen et al (2019) studied the M8.7 CRF associated with a type IV radio burst on 2014 December 17. The microwave sources at 1.2 2.0 GHz are just located at the flare site where null-point reconnection occurs. For 17 GHz, the microwave source is close to the two ribbons. For 34 GHz, there are one loop-top and two footpoint sources. In a follow-up work, Lee et al (2020) reanalyzed the same flare in detail and provided new evidence for the breakout-type eruption in a fan-spine structure (Wyper et al, 2017). Using high-cadence observation with the Nançay Radioheliograph (NRH; Ker- draon and Delouis, 1997), Duan et al (2022) obtained the trajectory of the microwave source at 150 MHz during a C1.4 class, jet-related CRF. Moving along closed magnetic field (outer spine), the radio source is produced by the accelerated electrons during null-point reconnection. Besides, the nonthermal

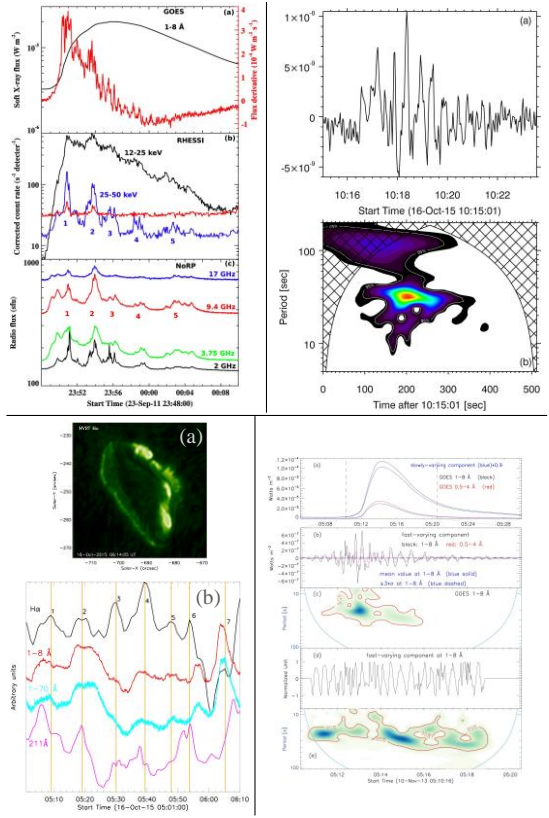

**Fig. 3** From top-left to bottom-right: Four examples of QPPs in CRFs (credit: Kumar et al (2016b), Zhang et al (2016a), Li et al (2020b), Ning et al (2022)).

electrons propagating upward along open magnetic field produce a concurrent type III burst.

# 4 Energy transport and partition

In the standard two-ribbon flare model, the flare-accelerated electrons stream down along the reconnected field lines and knock into the dense chromosphere. Impulsive precipitation of energy causes significant plasma heating to $\sim 10^7$ K and emissions in H$\alpha$, UV, EUV, HXR, and radio wavelengths at the flare ribbons (Brown, 1971). The hot plasmas flow up and fill the post-flare loops at speeds of a few hundred km s$^{-1}$, known as chromospheric evaporation (Acton et al, 1982; Antonucci et al, 1982; Fisher et al, 1985a,b). At the same time, the localized cool and dense plasmas flow down as a result of conservation of momentum, known as chromospheric condensation. Apart from nonthermal electrons, thermal conduction is capable of driving chromospheric evaporation (Ashfield and Longcope, 2021). The upflows (downflows) observed in high-temperature (low-temperature) spectral lines are characterized by blueshifts (redshifts), respectively. Recently, during the X1.3 flare on 2022 March 30 observed with IRIS, Xu et al (2023) discovered downward motion with extraordinarily high speeds (85 160 km s⁻¹) in UV lines (Mg II, C II, and Si IV). There are two types of chromospheric evaporation according to the injected energy flux density. Strong upflows at speeds of a few hundred km s⁻¹ and downflows at speeds of tens of km s⁻¹ are simultaneously observed during explosive evaporations (Brosius and Phillips, 2004; Milligan and Dennis, 2009; Li et al, 2015b; Tian et al, 2015). Moderate upflows are simultaneously observed in spectral lines formed in coronal and transition regions during gentle evaporations (Battaglia et al, 2009; Sadykov et al, 2015). Different types of evaporation may occur at different stages of flare evolution (Li et al, 2015d). Imaging observations of fast converging flows from the footpoints to the top of flare loops in EUV and HXR wavelengths have been reported (Liu et al, 2006; Ning et al, 2009; Li et al, 2017b).

Using high-resolution observations of the Interface Region Imaging Spectrograph (IRIS; De Pontieu et al, 2014) spacecraft, Zhang et al (2016b) studied a C4.2 class, jet-related CRF as a result of mini-filament eruption on 2015 October 16. Concurrent upflow (35 120 km s⁻¹) in the Fe XXI 1354.09 Å line and downflow (10 60 km s⁻¹) in the Si IV 1393.77 Å line are detected on the inner and circular flare ribbons (Figure 4). It is concluded that the explosive chromospheric evaporation is driven by nonthermal electrons accelerated by null-point magnetic reconnection.

Subsequently, Zhang et al (2016a) investigated another C-class CRF in the same AR, which was ∼ 3 hrs earlier than the C4.2 flare. Since the brightest part of the flare ribbons was missed by the IRIS raster, there was no response in the Fe XXI 1354.09 Å line. However, the Si IV 1402.77 Å line shows periodic redshifts, indicating periodic downflow in the chromosphere (Tian and Chen, 2018). The downflow speed is tightly correlated with the Si iv line intensity as well as the SXR flux derivative, which is a convincing signature of explosive chromospheric evaporation (Li et al, 2023a).

Apart from spectroscopic observations, direct imaging observation of chromospheric evaporation in CRF has been explored (Zhang et al, 2019b). During the impulsive phase of the C5.5 flare on 2014 August 24, converging motion and filling process in the flare loop are detected in AIA 131 Å and two XRT filters. The evaporation lasts for 6 minutes with an average speed of 170 km s⁻¹. Using the observed values of electron number density ( $1.8 \times 10^{10}$ cm⁻³), loop width (∼8"), and loop length (∼$5.7 \times 10^9$ cm), about $4.5 \times 10^{13}$ g chromospheric materials are evaporated into the single flare loop. Assuming that the CRF consists of 5 loops, a total amount of $2.3 \times 10^{14}$ g plasmas flow into the corona with thermal energy of $5.5 \times 10^{29}$ erg.

White-light flares (WLFs) are a special kind of flares with an enhanced emission at the visible continuum originating from the lower chromosphere down to the middle photosphere (Carrington, 1859; Hudson, 1972). The energy transport mechanisms include electron beam precipitation, heat conduction,

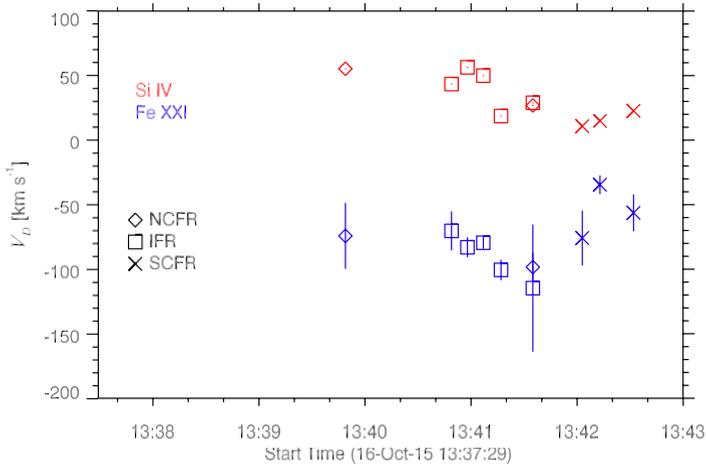

**Fig. 4** Time evolutions of the Doppler velocities at the inner ribbon (squares) and circular ribbon (diamonds and crosses) of the C4.2 CRF on 2015 October 16 (credit: Zhang et al (2016b)).

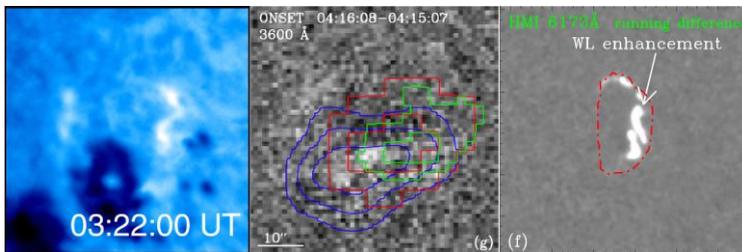

**Fig. 5** From left to right: three examples of circular WLFs on 2015 March 10 (credit: Hao et al (2017)), 2012 May 10 (credit: Song et al (2018)), and 2013 November 5 (credit: Song and Tian (2018)).

Alfvén waves, and radiative backwarming (Song et al, 2023). Hao et al (2017) analyzed the M5.1 circular WLF observed by SDO/AIA and the Optical and Near-infrared Solar Eruption Tracer (ONSET; Fang et al, 2013) on 2015 March 10 (see Figure 5). The bright kernels stand out in 3600 Å, 4250 Å, and Hα line center during the impulsive phase, with the maximal relative enhancement reaching ∼0.3. The origins of those WL kernels are discussed. Song et al (2018) investigated another M-class WLF on 2012 May 10 (see Figure 5), finding a close spatial and temporal relationship between the WL, HXR, and microwave emissions, which is accordant with the characteristic of type I WLF (Fang and Ding, 1995). The analysis favors the backwarming mechanism for the WL emission. To figure out the percentages of WLFs, Song and Tian (2018) selected 90 CRFs, including 8 X-class, 34 M-class, and 48 C- and B-class flares. It is found that the occurrence rate of WLFs increase with flare class, and there is a positive correlation between the WL enhancement and flare class for X-class WLFs.

The global energetics of two-ribbon flares have been extensively investigated using multiwavelength observations. Emslie et al (2004) studied the energy partitions of two X-class eruptive flares in 2002. The orders of magnitude of energy components are $\sim 10^{31}$, $10^{32}$, and $10^{32}$ erg for flares, CMEs, and magnetic fields, respectively. In a follow-up work, Emslie et al (2005) improved the energy estimate for flare thermal energies and added the calculations of total radiant energy, finding that the flare and CME have comparable energies ($10^{32}$ erg). Feng et al (2013) calculated the different energy com- ponents of the X2.1 flare on 2011 September 6. The 3D reconstruction of the CME morphology using the mask fitting method (Feng et al, 2012) is utilized to get a more precise estimation of the CME energy. It is concluded that the flare and CME consume a similar amount of magnetic free energy. Li et al (2023b) exquisitely calculated energy partitions of the X9.3 flare, related CME, and solar energetic particles (SEPs) on 2017 September 6. It is revealed that the prompt component of the SEP event contains a negligible fraction of the flare energy.

Energetics of CRFs, however, has rarely been explored. For the first time, Zhang et al (2019a) studied two homologous M1.1 CRFs originating from AR 12434 without CMEs. Various energy components are calculated, including the peak thermal energy, nonthermal energy of electrons, total radiative loss of hot plasma, and radiant energies of the flares. Interestingly, the two flares present similar morphology and energetics. The ratios of energy components between the two flares have marginal fluctuation. In a follow-up study, Cai et al (2021) investigated energy partitions in four confined CRFs and found that the values of energy components increase systematically with flare class. Moreover, the ratio of nonthermal energy to magnetic free energy ($\frac{E_{\mathrm{nth}}}{E_{\mathrm{mag}}}$) may play a critical role in distinguishing confined from eruptive flares, since there is no consumption of magnetic free energy by CMEs in confined eruptions. This conjecture requires further statistical analysis (Li et al, 2020e).

# 5 Related activities

## 5.1 Remote brightenings or ribbons

As mentioned above, two-ribbon flares present contraction, separation, and elongation in the ribbons observed in UV, EUV, and H$\alpha$ wavelengths (Qiu et al, 2002; Ji et al, 2004, 2006; Qiu et al, 2017). Multiple ribbons, including remote brightenings or ribbons, are not so common (e.g., Wang et al, 2014; Qiu et al, 2020; Joshi et al, 2022). On the contrary, CRFs are frequently associated with remote brightenings (Sun et al, 2013; Kumar et al, 2016b; Masson et al, 2017; Chen et al, 2019; Devi et al, 2020; Yang et al, 2020b). Considering that the magnetic topology of a CRF is characterized by a null point, a dome- like fan surface, an inner spine, and an outer spine, the accelerated electrons may propagate to the remote site along the outer spine and generate remote brightenings or ribbons (Wang and Liu, 2012, see also Figure 1). As a consequence, the time lags between remote brightenings and main ribbons are a few seconds. Hernandez-Perez et al (2017) analyzed an M2.1 CRF on 2015 January 29. Two primary ribbons are located under the fan surface. Secondary brightenings are ∼108 Mm away from the primary ribbons. The heating mechanism of the secondary brightenings is dissipation of kinetic energy of the plasma flow originating from the flare site, rather than non-thermal electrons accelerated by magnetic reconnection. Figure 6 shows four examples of remote brightenings or ribbons related to circular ribbons.

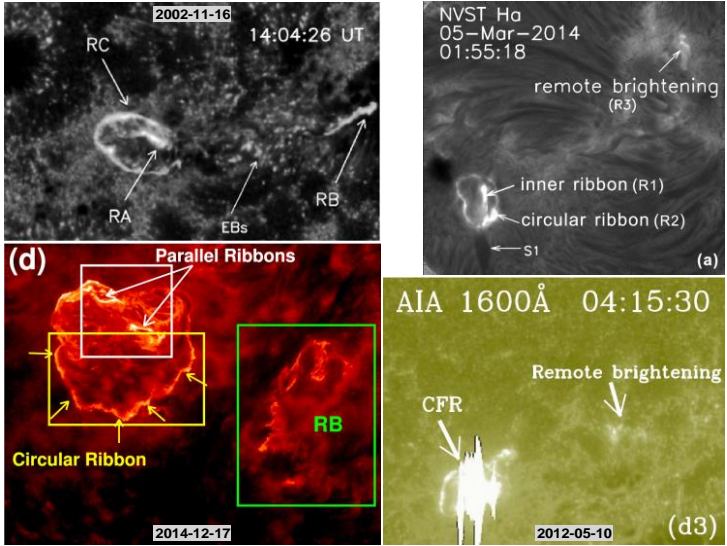

**Fig. 6** Four examples of remote brightenings or ribbons related to circular ribbons (credit: Masson et al (2009); Xu et al (2017); Joshi et al (2021); Song et al (2018)).

Among the 134 events, 76 (∼57%) are accompanied by remote brightenings or ribbons, implying that a larger number of the outer spines are closed, i.e. connecting to remote sites (Zhang et al, 2022d). The total length ($L_{RB}$) and average distance ($D_{RB}$) of the brightening from the center of the circular ribbon are separately measured. $L_{RB}$ lies in the range of 6.2 and 381 Mm, with an average value of ∼84 Mm. $D_{RB}$ lies in the range of 28 and 186 Mm, with an average value of ∼89 Mm. A positive correlation between the two parameters is uncovered with a correlation coefficient of ∼0.65. A linear fitting results in $D_{RB} = 67.56 + 0.26L_{RB}$, suggesting that longer remote ribbons tend to be further from the circular ribbons. The aspect ratio $\kappa = \frac{D_{RB}}{2r_{CF}}$ is between 0.8 and 6.7 with an average value of 2.9, which is close to the values in 3D numerical simulations (Wyper and DeVore, 2016; Wyper et al, 2016). Despite that remote brightenings or ribbons are widespread, their physical properties have rarely been investigated, such as chromospheric evaporation and QPP induced by nonthermal electrons (Zhang and Ji, 2013; Zhang et al, 2016a).

## 5.2 EUV late phases

Generally speaking, as the hot plasmas ( $\sim 10^7$ K) in post-flare loops (PFLs) cool down due to heat conduction and radiative loss (Cargill, 1994; Cargill et al, 1995), the EUV emission reaches a peak successively from high to low temperatures. For the first time, Woods et al (2011) discovered a second enhancement using the EUV irradiance observations from the EUV Variability Experiment (EVE; Woods et al, 2012) on board SDO. Different from the main peak, the second peak in the warm emissions (e.g., Fe xv 284 Å, Fe xvi 335 Å) is called "EUV late phase". Time lag between the late-phase peak and the SXR peak ranges from tens of minutes to a few hours (Dai et al, 2013, 2018; Li et al, 2014). The EUV late phase could be lower, equal to, or even higher than the first peak (Liu et al, 2015b; Zhou et al, 2019).

The first and second peaks in EUV emissions are believed to originate from different sets of flare loops. The loops creating the late-phase emissions are usually longer or higher than the main flare loops (Liu et al, 2013b; Li et al, 2014; Zhang et al, 2021a). However, the mechanism of EUV late phase is still under debate. Two candidates are proposed to account for the late-phase emission: additional heating (Dai et al, 2013; Zhou et al, 2019; Zhang et al, 2022a) and long-lasting cooling of the late-phase loops (Liu et al, 2013b; Li et al, 2014; Masson et al, 2017). Sun et al (2013) studied the EUV late phase related to the M2 class CRF on 2011 November 15 (fourth row in Figure 7) and concluded that the late phase results from cooling of large-scale post-reconnection loops, although additional heating may also be needed. Li et al (2014) reanalyzed some of the CRFs showing EUV late phase and found that the late phase loops are either hot spine field lines related to a null point or large-scale loops of multipolar magnetic fields. Chen et al (2023b) investigated the X1.8 flare as a result of nonradial eruption of a magnetic flux rope from a fan-spine structure on 2011 September 7. Atypical plateau-like EUV late phase rather than a single peak is discovered, which is explained by a combination of emissions from a group of late-phase loops with different lengths and cooling times. Chen et al (2020) investigated 55 large flares with EUV late phases, 19 ($\sim$35%) of which are CRFs. Besides, only 22 (40%) of them are eruptive while 33 (60%) of them are confined. Figure 7 show five representative CRFs with EUV late phases, featuring the late-phase loops, flare ribbons, and 3D magnetic configurations.

## 5.3 Coronal jets and type III radio bursts

Coronal jets are hot and collimated plasmas flowing along open or large-scale closed coronal loops (see Raouafi et al, 2016; Shen, 2021, and references therein). Using observations from the Soft X-ray Telescope (SXT; Tsuneta et al, 1991) on board Yohkoh, the basic properties of X-ray jets are derived (Shimojo et al, 1996). The hot jets observed in EUV or X-ray are mostly accompanied by cool surges (Chae et al, 1999; Mulay et al, 2016; Li et al, 2017c; Shen et al, 2017; Wang et al, 2023c). According to the morphology, coronal jets are usually classified into two types: the first type has an inverted-$Y$ shape resembling the famous Eiffel Tower (Shibata et al, 1992; Krucker et al, 2011b) and the second type has a two-sided extension (Jiang et al, 2013b; Shen et al, 2019a; Tan et al, 2022). It is widely accepted that jets are created by magnetic reconnection between the emerging magnetic field and previously existing open field with opposite polarity (Shibata et al, 1992; Yokoyama and Shibata, 1995; Moreno-Insertis et al, 2008; Baumann and Nordlund, 2012; Ni et al, 2017; Li et al, 2023c) or magnetic flux cancelation (Panesar et al, 2016, 2017; Sterling et al, 2017). Minifilament eruptions are also considered as an important mechanism of jet formation (Sterling et al, 2015, 2016; Hong et al, 2016), especially for the blowout jets (Moore et al, 2010; Li et al, 2015c, 2017c). Pariat et al (2009) proposed a model for polar jets, in which twisting motion at the photosphere injects magnetic helicity and free energy into the upper atmosphere. An ideal MHD instability breaks the initial symmetry and triggers an impulsive release of energy via magnetic reconnection (top-left two panels of Figure 8). Such process will repeat when the twisting motion continues and generates homologous coronal jets (Pariat et al, 2010). Zhang et al (2012) studied two CBPs associated with jets observed by the X-Ray Telescope (XRT; Golub et al, 2007) on board the Hinode spacecraft (Kosugi et al, 2007) on 2007 March 16. The magnetic topology, featuring a null point, a fan dome, and spine lines, is excellently consistent with the above numerical model (top-right two panels of Figure 8). Nóbrega-Siverio et al (2023) demonstrated a 3D numerical model with a null point, which nicely explains the continuous Joule and viscous heating of CBPs for a few hours (Huang et al, 2012; Madjarska et al, 2012).

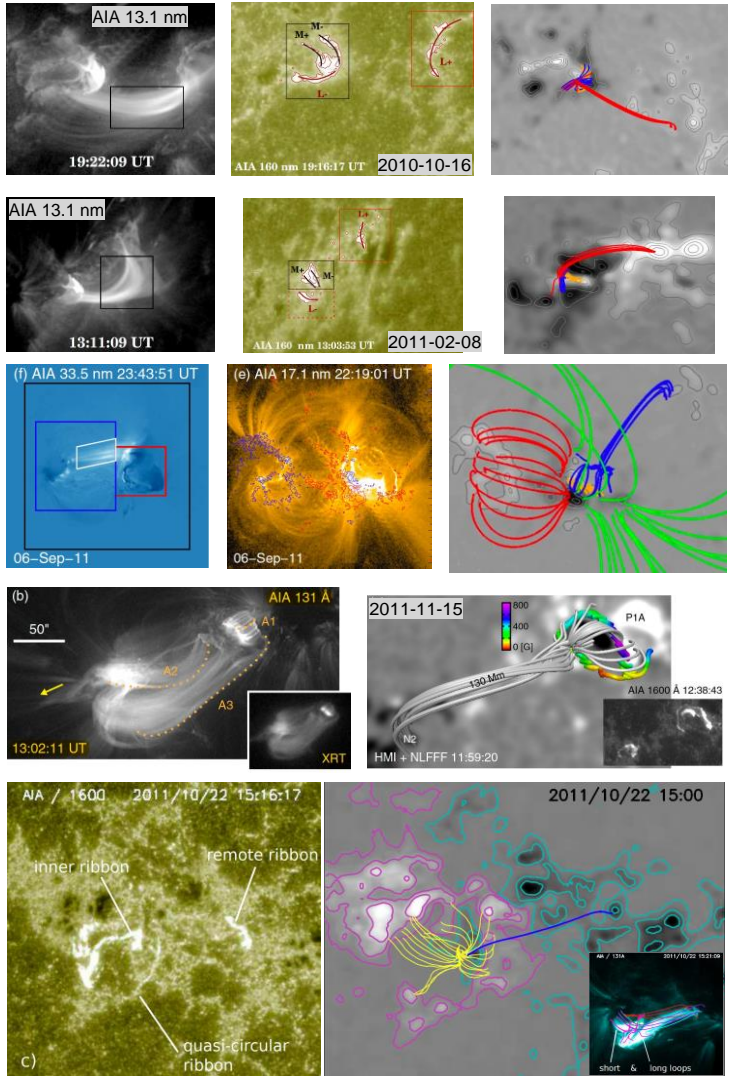

**Fig. 7** From top to bottom: five examples of CRFs with EUV late phases, including the flare loops, ribbons, and the 3D magnetic configurations (credit: Liu et al (2013b); Dai et al (2013); Li et al (2014); Sun et al (2013); Masson et al (2017)).

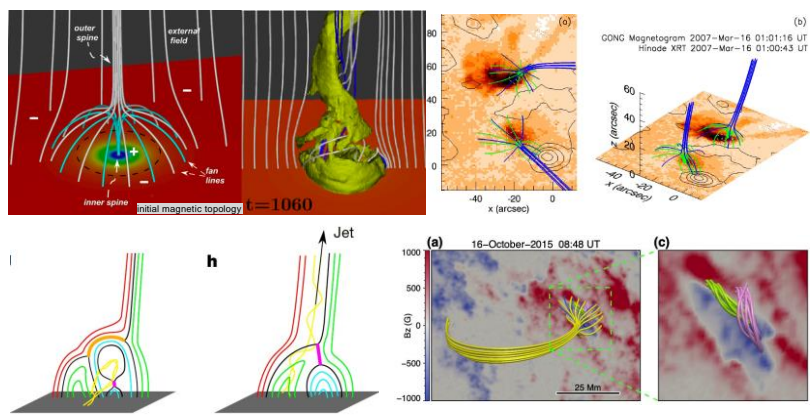

**Fig. 8** Top-left two panels: 3D magnetic configuration of the jet at $t = 0$ and $t = 1060$ in numerical simulation (credit: Pariat et al (2009)). Top-right two panels: top view and perspective view of the 3D magnetic configuration of the jet-related CBP (credit: Zhang et al (2012)). Bottom-left two panels: schematic of breakout process in jets (credit: Wyper et al (2017)). Bottom-right two panels: top view of the 3D magnetic configuration of the jet-related CRF and the underlying mini-filaments (credit: Zhang et al (2021b)).

Inspired by the famous breakout model for CMEs (Antiochos et al, 1999; Lynch et al, 2004), Wyper et al (2017) put forward a universal model for solar eruptions. The breakout jet results from mini-filament eruption, during which magnetic reconnections successively take place at the breakout current sheet (BCS) near the null point and the flare current sheet (FCS) beneath the filament (bottom-left two panels of Figure 8). The kinetic energy of the jet increases and magnetic free energy decreases sharply when the flux rope opens and reconnects with the background field lines (Wyper et al, 2018). Observational evidences for this model are abundant. Kumar et al (2019a) reported direct observations of breakout reconnection in the fan-spine topology of an embedded bipole. Meanwhile, multiple small-scale plasmoids with bidirectional flows are observed as a result of quick reconnection in the BCS. The explosive jet with a filament eruption provides solid support for the breakout model. Plasmoids (or blobs) are frequently observed in coronal jets, which is explained by tearing-mode instability in current sheets (Zhang and Ji, 2014b; Wyper et al, 2016; Joshi et al, 2018; Zhang and Ni, 2019; Li and Yang, 2019; Joshi et al, 2020; Chen et al, 2022; Mandal et al, 2022; Cheng et al, 2023; Mulay et al, 2023; Yang et al, 2024). Kumar et al (2021a) studied the C2.4 CRF associated with a CME on 2015 April 20 and identified the features of erupting flux rope, FCS, and underlying flare arcade. Kumar et al (2019b) found that all the 27 equatorial coronal-hole jets are related to the fan-spine topology associated with CBPs, which are consistent with the breakout model of jets. Zhang et al (2021b) analyzed the C3.4 CRF associated with a coronal jet on 2015 October 16. The 3D magnetic topology of the flare is fan-spine structure (bottom-right two panels of Figure 8). The jet and flare result from eruption of a mini-filament. Kinetic evolution of the jet is divided into two phases: a slow rise phase $\backsim$ 131 km s⁻¹) and a fast-rise phase $\backsim$ 363 km s⁻¹). Magnetic reconnection at the breakout current sheet is characterized by a circular ribbon in the chromosphere, HXR source at the inner spine, and a type III radio burst. Spectroscopic observations from IRIS reveal significant bidirectional outflows, which are indicative of magnetic reconnection at the jet base (or FCS). The detailed analysis are strongly in favor of the breakout jet model (Wyper et al, 2018).

Type III radio bursts are frequently linked to solar flares when accelerated electron beams propagate outward along open magnetic field lines (Aschwanden et al, 1995; Mann et al, 1999; Morosan et al, 2014; Zhang and Ji, 2014a; Zhang et al, 2022c). For CRFs with a fan-spine structure, the electrons are accelerated by magnetic reconnection at the null point and propagate along the open outer spine to generate a type III burst (Zhang et al, 2016a; Zhang and Ni, 2019; Duan et al, 2022). The type III bursts are generally in correspondence with HXR peaks. Figure 9 shows eight examples of CRFs and their related type III bursts in the radio dynamic spectra.

In the 134 CRFs, 69 ($\sim$51%) and 63 ($\sim$47%) of them are accompanied by jets and type III radio bursts, respectively (Zhang et al, 2022d). There are no dependences of jet and type III burst production rates on flare magnitudes, indicating that magnetic configuration, instead of flare energy, is relevant to the generation of coronal jets and type III bursts.

## 5.4 CMEs, shock waves, and type II radio bursts

As mentioned in Section 2, a large number of CRFs originate from filament or flux rope eruptions, which might evolve into CMEs observed in WL coronagraphs. Figure 10 shows three examples of large CRFs observed by SDO/AIA and the related CMEs observed by the Large Angle Spectroscopic Coronagraph (LASCO; Brueckner et al, 1995) on board the Solar and Heliospheric Observatory (SOHO; Domingo et al, 1995) mission. All these flares are located in the southern hemisphere with GOES classes of X1.1, M8.7, and M6.9, respectively. All these CMEs are partial or full halo. Statistical investigation reveals that the occurrence rate of CMEs associated with CRFs is merely 28%, suggesting that most of CRFs are confined rather than eruptive events (Zhang et al, 2022d). Moreover, the CME speed is positively correlated with the peak SXR flux for these eruptive CRFs, indicating that larger flares are more likely to generate faster CMEs. Amari et al (2018) studied the confined X3.1 flare on 2014 October 24 and proposed the concept of "magnetic cage". A weaker magnetic cage is likely to produce a more powerful eruption with a CME. For CRFs usually associated with a dome-like fan-spine structure determined by a magnetic null point, it is conjectured that the dome-like fan surface serves as a solid cage to suppress a successful eruption, which may interpret the low occurrence rate of CMEs with CRFs (Wyper and DeVore, 2016; Yang and Zhang, 2018).

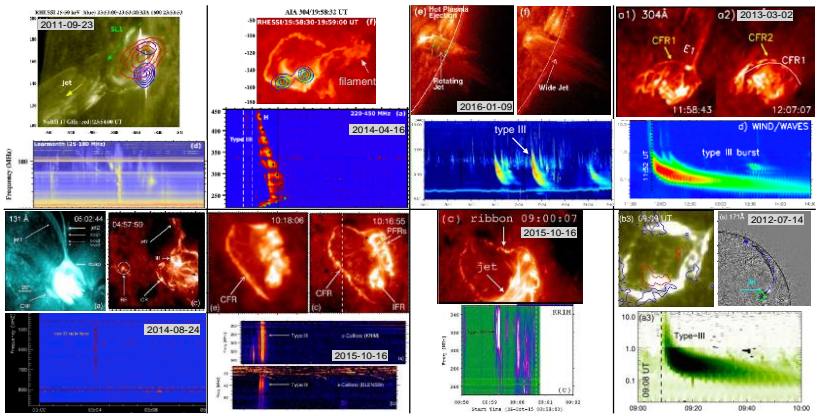

**Fig. 9** From top-left to bottom-right: eight examples of CRFs and the related type III radio bursts (credit: Kumar et al (2016a,b); Joshi et al (2018); Yang et al (2020a); Zhang and Ni (2019); Zhang et al (2016a, 2021b); Duan et al (2022)).

Fast CMEs are capable of driving shock waves, which are accompanied by type II radio bursts with drifting frequencies (Morosan et al, 2019; Zhang et al, 2022b). Figure 11 shows two examples of CRFs observed by SDO/AIA, the CME-driven shock waves observed by LASCO, and the related type II bursts observed by ground-based radio telescopes (Zhang et al, 2022d).

## 5.5 Coronal dimmings

Coronal dimmings are dark regions of greatly reduced emission in EUV and SXR (Hudson et al, 1996; Moses et al, 1997; Thompson et al, 1998; Delannée and Aulanier, 1999). It is generally accepted that the dark voids with enhanced density depletion are intimately associated with the eruptions of CMEs (Harrison and Lyons, 2000; Veronig et al, 2019; Chikunova et al, 2023). The dimming region is characterized by massive blueshifts at speeds of ∼100 km s$^{-1}$ in EUV spectroscopic observations, so that the mass loss could be used to estimate the mass of related CME (Harra and Sterling, 2001; Jin et al, 2009). Coronal dimmings are classified into core dimmings and secondary dimmings (Zarro et al, 1999). Core dimmings are cospatial with the footpoints of erupting filaments or flux ropes, while secondary dimmings are due to quick expansion of the overlying magnetic structures. Ahead of the spreading coronal dimmings, there is always an EUV wave (Chen et al, 2002, 2005; Patsourakos et al, 2010; Ma et al, 2011; Liu and Ofman, 2014; Dai et al, 2023).

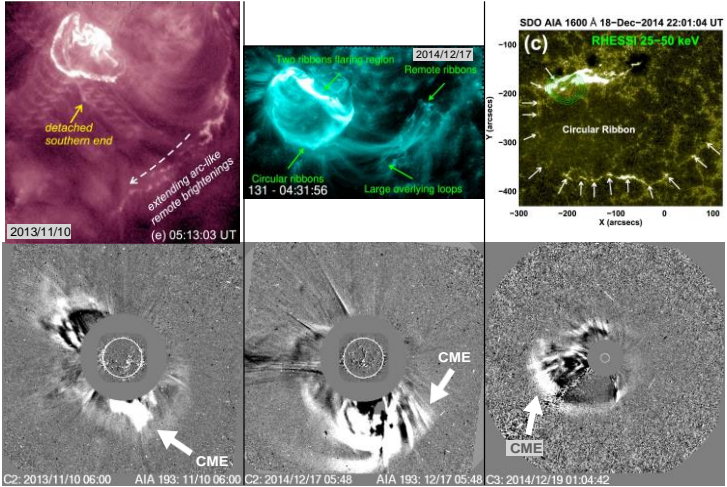

**Fig. 10** Three examples of CRFs observed by SDO/AIA and the related CMEs observed by SOHO/LASCO on 2013 November 10 (left panels), 2014 December 17 (middle panels), and 2014 December 18 (right panels) (credit: Liu et al (2020); Chen et al (2019); Joshi et al (2017)).

Coronal dimmings usually appear in the acceleration phase of a CME, which is concurrent with the impulsive phase of the related flare (Zhang et al, 2001). For the first time, Zhang et al (2017) detected pre-flare coronal dimmings near the conjugate footpoints of a flux rope, which started 96 minutes before the X1.6 flare on 2014 September 10. Such pre-eruption dimmings are further validated by multiwavelength observations and are important for space weather prediction (Qiu and Cheng, 2017; Wang et al, 2019, 2023a; Prasad et al, 2020; Pan et al, 2022). Kumar et al (2021a) investigated two events of pseudostreamer jets and CMEs resulting from filament eruptions on 2015 April 20 and 21. Pre-eruption opening or dimming are detected as a result of transverse motion away from the initial null point location prior to the flare onset (see Figure 12).

For CRFs associated with CMEs, localized or post-eruption dimmings as a result of density decrease at the flare site are formed after the eruption of filaments or flux ropes (see Figure 12). Zhang and Zheng (2020) studied the M1.1 CRF in AR 12434 on 2015 October 16. The SXR flux of the flare increased rapidly from ∼06:11 UT to the peak at ∼06:16 UT. Owing to the Neupert effect (Neupert, 1968), the HXR peak time (06:13:48 UT) was a little earlier. Several minutes before the HXR peak time, small-scale, weak pre-flare dimming showed up 180 Mm away from the flare site in the southwest direction. Following the pre-flare dimming observed merely in 131 and 171 Å, long and narrow dimmings appeared simultaneously in all AIA EUV passbands except 304 Å, indicating that the dimmings are caused by density decrease rather than temperature variation (see Figure 13). The large-scale dimmings extended in the southeast direction, leading to an increasing area up to $1.2 \times 10^4$ Mm$^2$ in 193 Å. The relative intensity decrease reached 80%-90%, followed by a long, gradual recovery. Considering that the remote dimmings are connected with the flare site by large-scale coronal loops, the dimmings are most probably caused by density depletion of these loops as a result of field-line stretching during the filament eruption. Interestingly, during the C3.4 homologous CRF that occurred nearly three hours later, remote dimmings were also detected at the same place (Zhang, 2020, see their Fig. 6). Note that remote dimmings are different from the widespread secondary dimmings originating from the flare sites and expanding almost in all directions (Dissauer et al, 2018).

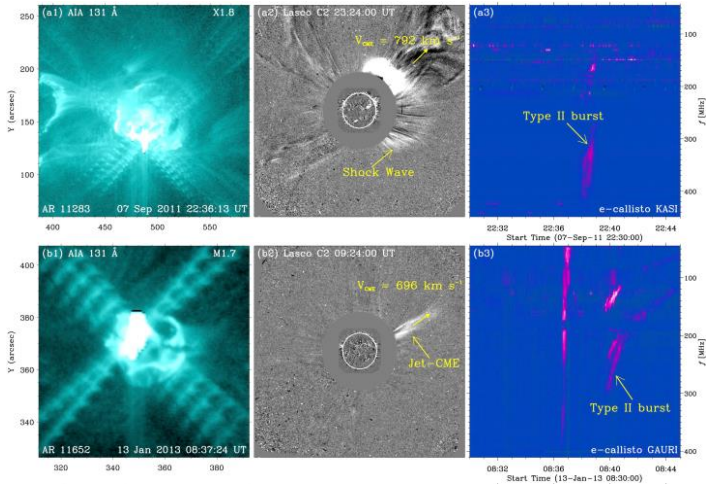

**Fig. 11** Two examples of CRFs observed by SDO/AIA, the CME-driven shock waves observed by SOHO/LASCO, and the related type II bursts observed by ground-based radio telescopes. (credit: Zhang et al (2022d)).

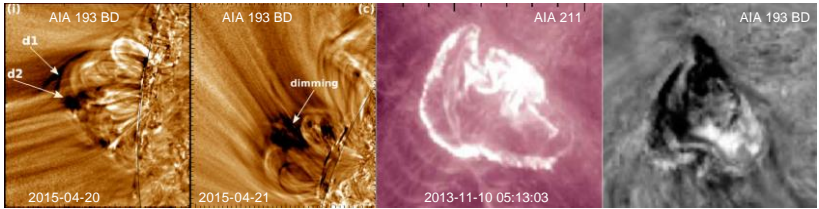

Fig. 12 Left two panels: Pre-eruption dimmings associated with CRFs on 2015 April 20 and 21 (credit: Kumar et al (2021a)). Right two panels: Localized or post-eruption dimming associated with the CRF on 2013 November 10 (credit: Liu et al (2020)).

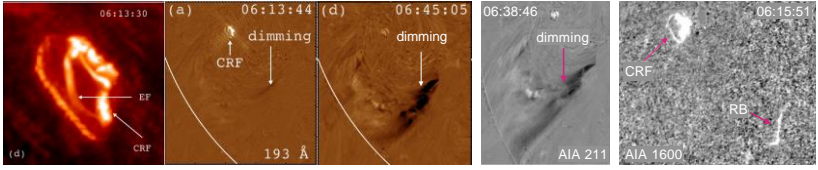

**Fig. 13** From left to right: erupting mini-filament that leads to the CRF observed by AIA 304 Å image on 2015 October 16, pre-flare dimming and featherlike remote dimmings associated with the flare observed by AIA 193 Å and 211 Å images, and remote brightenings adjacent to the dimmings observed by AIA 1600 Å image (credit: Zhang and Zheng (2020)).

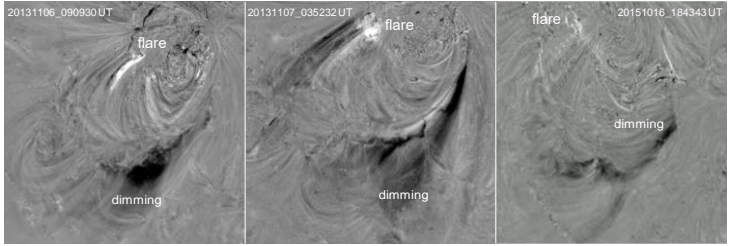

**Fig. 14** Three examples of large-scale, remote dimmings related to CRFs observed by AIA 193 Å (base-difference images).

Liu et al (2020) investigated the X1.1 CRF on 2013 November 10, which resulted from a flux rope eruption beneath the fan dome. Initial remote brightening appeared to the southwest of flare site. Extended, arc-like remote brightenings propagated in the southeast direction from the initial location of brightening, reaching a splendid scale of ∼400″ (∼290 Mm). Meanwhile, the remote brightenings are accompanied by coronal dimmings. The coexisting remote brightenings and dimmings are caused by opening of the outer spine-like loops during the fast eruption of the flux rope in a whipping fashion. Figure 14 shows three examples of large-scale, remote dimmings related to CRFs on 2013 November 6, 2013 November 7, and 2015 October 16, respectively.

Prasad et al (2020) studied the X2.1 flare on 2011 September 6 in AR 11283 and identified three patches of coronal dimmings: a ring-shaped dimming close to the main flare site, a circular dimming east to the flare site, and a remote dimming northwest to the flare site (see their Fig. 2). The footpoints of the dome surface of the null point are cospatial with the ring-shaped dimming. The two dimmings far from the flare site are cospatial with the remote flare ribbons, and their origin are still unclear.

## 5.6 Coronal loop and filament oscillations

MHD waves and oscillations are ubiquitous in the solar atmosphere, including the sunspots (Khomenko and Collados, 2015; Yuan et al, 2023), chromosphere (De Pontieu et al, 2007; Jess et al, 2015), corona (Li et al, 2020a; Nakariakov and Kolotkov, 2020; Banerjee et al, 2021), and solar flares (Tian et al, 2016; Li et al, 2017a). These waves are classified into three modes, fast-mode waves (Nakariakov et al, 2021), slow-mode waves (Wang et al, 2021), and Alfvén waves (Erdélyi and Fedun, 2007; Tomczyk et al, 2007; Jess et al, 2009; McIntosh et al, 2011). The dissipation of wave energy plays an essential role in coronal heating (Antolin and Shibata, 2010; van Ballegooijen et al, 2011; Arregui, 2015; Cranmer et al, 2015; Van Doorsselaere et al, 2020; Srivastava et al, 2021; Lim et al, 2023). Kink oscillations were first discovered with high-resolution observations of TRACE in 171 Å (Aschwanden et al, 1999; Nakariakov et al, 1999; Van Doorsselaere et al, 2008a). The observed periods and loop lengths are utilized to carry out diagnostics of magnetic field strength and Alfvén speed of oscillating loops (Nakariakov and Ofman, 2001; Aschwanden et al, 2002; Van Doorsselaere et al, 2008b). Using high-resolution observations with the Coronal Multi-channel Polarimeter (CoMP; Tomczyk et al, 2008), Yang et al (2020c) obtained the map of the plane-of-sky component of the global coronal magnetic field between 1.05 $R_0$ and 1.35 $R_0$.

The amplitudes of kink oscillations usually damp with time due to phase mixing and resonant absorption (Goossens et al, 2002; Ruderman and Roberts, 2002; Terradas et al, 2006; Goddard and Nakariakov, 2016; Pascoe et al, 2016a,b). Kink oscillations last for a few cycles before fading out and the quality factor ($q = \frac{\tau}{P}$) is lower than 10 in most cases. Recently, decayless oscillations of coronal loops become a topic of great interest (Nisticò et al, 2013; Mandal et al, 2021; Gao et al, 2022; Zhong et al, 2022, 2023; Li and Long, 2023), especially using observations of the Extreme Ultraviolet Imager (EUI; Rochus et al, 2020) on board the Solar Orbiter (Müller et al, 2020). Compared with decaying oscillations, decayless oscillations have much shorter amplitudes and longer durations.

Kumar et al (2016a) studied an EUV wave and related type II radio burst, which were caused by a confined flare with fan-spine topology on 2014 April 16. Transverse oscillations are excited in many loops within the arcade when the fast EUV wave passes through the loops. The period and damping time of kink oscillation of one loop are ~210 s and ~506 s, respectively. The kink speed and Alfvén speed of the loop are estimated to be ~1140 and ~840 km s$^{-1}$. On 2014 March 5, a series of homologous flares occurred in AR 11991 (Xu et al, 2017). Zhang et al (2020a) studied the C2.8 class flare (CRF1) and the M1.0 class flare (CRF2) with an interval of ~14 minutes. CRF1 excited decayless transverse oscillation of the coronal loop ($L \approx 130$ Mm) with amplitudes of 310–510 km. On the contrary, CRF2 excited decaying loop oscillation of the same loop with greatly enhanced amplitudes of 1250–1280 km (see Figure 15). The periods (<120 s) of the loop oscillations are comparable during the two phases. Hence, the observations provide new evidence of significantly amplified amplitudes of oscillations by successive drivers (Nisticò et al, 2013).

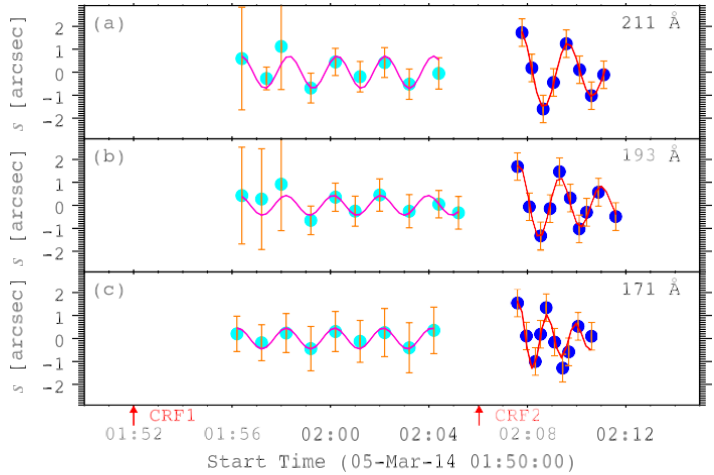

**Fig. 15** Decayless loop oscillation triggered by CRF1 (cyan circles) and decaying loop oscillation triggered by CRF2 (blue circles) on 2014 March 5 (credit: Zhang et al (2020a)).

Similarly, filament oscillations with enhanced amplitudes are detected. Zhang et al (2020b) investigated large-amplitude, longitudinal filament oscillations triggered by two homologous flares with an interval of $\sim$ 4 hr on 2010 October 18. The amplitudes of filament oscillations excited by the C2.6 eruptive flare with a jet-like CME were markedly larger than the amplitudes of oscillation excited by the C1.3 confined flare. The observations are satisfactorily reproduced by one-dimensional, hydrodynamic numerical simulation using the MPI-AMRVAC code (Porth et al, 2014; Keppens et al, 2023). Kink oscillations of the large-scale coronal loops ($L_{\approx}$ 370 Mm) excited by coronal jets related to CRFs in AR 12434 are explored by Zhang (2020) and Dai et al (2021). Interestingly, transverse filament oscillation is simultaneously detected $\sim$ 140 Mm away from the flare site, which is probably induced by secondary magnetic reconnection near the filament (Zhang, 2020).

# 6 Discussion

Owing to the big difference between the magnetic topology of two-ribbon flares and CRFs, they have distinct observational characteristics in many aspects. Firstly, a two-ribbon flare has a pair of conjugate ribbons in the chromosphere and transition region (Dennis and Pernak, 2009; Jing et al, 2016). A CRF consists of an inner and a circular ribbon at least. A remote ribbon or brightening is detected when the outer spine is linked to a remote polarity on the solar surface (Section 5.1). Occasionally, a pair of ribbons could be identified under the fan dome when a minifilament or flux rope erupts (Joshi et al, 2015, 2017). In consequence, CRFs tend to show multiple ribbons. Secondly, contraction and separation motions are frequently observed in two-ribbon flares (Ji et al, 2006; Liu et al, 2013c). However, such motions are rarely observed and reported due to the stability and persistence of the fan-spine structure. Elongation motion as a result of sequential reconnection have been observed in both types of flares (Qiu et al, 2017; Li et al, 2018a). Thirdly, magnetic reconnection takes place in the current sheets beneath the erupting flux ropes in most theoretical models (Shibata et al, 1995; Chen and Shibata, 2000; Lin and Forbes, 2000; Moore et al, 2001). In the breakout model, magnetic reconnection initially happens at the X-type null point above the sheared arcades and at the current sheet behind the flux rope after its eruption (Antiochos et al, 1999; Lynch et al, 2004). Sometimes, reconnection with the overlying magnetic field may lead to erosion of a flux rope and a confined eruption (Chen et al, 2023a). For CRFs, magnetic reconnection and particle acceleration take place mainly at the 3D null points, including spine reconnection and fan reconnection (Masson et al, 2009; Priest and Pontin, 2009; Pontin et al, 2013). In addition, reconnection might be detected under the erupting flux rope (Joshi et al, 2015; Song et al, 2018). Considering the ubiquity of QSLs, magnetic reconnections at QSLs are found in both kinds of flares (Masson et al, 2009; Aulanier et al, 2012, 2013; Janvier et al, 2013; Cheng and Ding, 2016). Finally, the fan-spine structure, especially the outer spine, allows for remote ribbons and EUV late phases associated with CRFs (Sections 5.1 and 5.2), which are scarcely observed in two-ribbon flares.

Despite of huge advances in observational and theoretical investigations on CRFs in the past years, there are still a few open questions, which need to be addressed in the future. The most important problem, in my opinion, is particle acceleration in CRFs. Although electric field plays a critical role in accelerating particles around 3D null points (Section 3), other mechanisms may play a part, such as termination shocks created by the reconnection outflow (Tsuneta and Naito, 1998; Mann et al, 2009; Chen et al, 2015; Takasao and Shibata, 2016; Kong et al, 2019), turbulence above the flare loop top (Petrosian and Liu, 2004; Kontar et al, 2017), contracting magnetic islands (Drake et al, 2006; Dahlin et al, 2014), and multi-island coalescence (Oka et al, 2010; Ni et al, 2015; Arnold et al, 2021; Li et al, 2022b) in current sheets. Coordinated, high-resolution, and multiwavelength observations with the NVST, Goode Solar Telescope (GST; Cao et al, 2010), the Expanded Owens Valley Solar Array (EOVSA; Nita et al, 2016), the Broadband Solar Radio Dynamic Spectrometer (Shang et al, 2022, 2023), the Spectrometer/Telescope for Imaging X-rays (STIX; Krucker et al, 2020) on board the Solar Orbiter, the Chinese Hα Solar Explorer (CHASE; Li et al, 2022a), and the Hard X-ray Imager (HXI; Su et al, 2019) on board the Advanced Space-based Solar Observatory (ASO-S; Gan et al, 2019), are extremely desired to shed light on this issue. The second problem is the cause of QPPs in CRFs. As mentioned in Section 3, QPPs in CRFs are often observed in multiwavelengths, such as HXR, radio, and UV (Zhang et al, 2016a). The detected periods may vary at different locations and different phases of flares (Chen et al, 2019), suggesting that different mechanisms may account for the complex behavior. Finally, the true morphologies of CRFs and their HXR sources based on 3D reconstruction from multiview observations have never been explored and reported (Krucker et al, 2010). EUV images obtained from SDO/AIA, the Solar Ultraviolet Imager (SUVI; Seaton and Darnel, 2018) on board the GOES-16 spacecrafts, and the EUI on board Solar Orbiter may enable us to perform such reconstructions and compare with data-constrained or data-driven MHD simulations (Jiang et al, 2016; Guo et al, 2023b,a).

# 7 Summary

Owing to the prevalence of 3D magnetic null points and the corresponding fan-spine topology in the solar atmosphere, CRFs are regularly observed in UV, EUV, and Hα passbands. Spine reconnection and fan reconnection around the null points are predominantly responsible for the energy release and subsequent particle acceleration. Slipping reconnection at QSLs may explain the sequential brightening or rapid degradation of the circular ribbons. Periodic or quasi-periodic acceleration and precipitation of nonthermal particles in the chromosphere produce observed QPPs of CRFs in multiple altitudes as well as wavelengths. Like two-ribbon flares, the injected high-energy particles result in explosive evaporation in circular and inner ribbons, which is characterized by simultaneous blueshifts in the coronal emission lines and redshifts in the chromospheric emission lines. Homologous CRFs residing in the same AR present similar morphology, evolution, and energy partition.

The peculiar topology of CRFs with closed outer spines facilitates remote brightenings and EUV late phases, which are uncommon in two-ribbon flares. Besides, CRFs are often accompanied by coronal jets, type III radio bursts, CMEs, shock waves, coronal dimmings, and kink oscillations in coronal loops and filaments. MHD numerical simulations are very helpful to decipher the key problems that are still unclear up to now. With the peak year of $25^{th}$ solar cycle approaching, splendid explosions, including jets, flares, and CMEs, are blooming. Multiwavelength and multipoint observations with state-of-the-art instruments are enormously desired to make a breakthrough. The findings in CRFs are important for a comprehensive understanding of    solar flares and have implication for stellar flares.

**Supplementary information.** If your article has accompanying supplementary file/s please state so here.

Authors reporting data from electrophoretic gels and blots should supply the full unprocessed scans for key as part of their Supplementary information. This may be requested by the editorial team/s if it is missing.

Please refer to Journal-level guidance for any specific requirements.

Acknowledgments. The author appreciates the reviewers for valuable suggestions to improve the quality of this review. The author would like to express his gratitude to Profs. Pengfei Chen, Hui Tian, and Yuandeng Shen for helpful discussions. This work is supported by the Strategic Priority Research Program of the Chinese Academy of Sciences (CAS), Grant No. XDB0560000, NSFC under the grant number 12373065, the National Key R&D Program of China 2021YFA1600500 (2021YFA1600502), the Surface Project of Jiangsu Province under the number BK20231510, and Yunnan Key Laboratory of Solar Physics and Space Science under the number YNSPCC202206.

# Declarations

**Conflict of interest** The author claims that he has no conflicts of interest.

Some journals require declarations to be submitted in a standardised format. Please check the Instructions for Authors of the journal to which you are submitting to see if you need to complete this section. If yes, your manuscript must contain the following sections under the heading 'Declarations':

- Funding
- Conflict of interest/Competing interests (check journal-specific guidelines for which heading to use)
- Ethics approval
- Consent to participate
- Consent for publication
- Availability of data and materials
- Code availability
- Authors' contributions

If any of the sections are not relevant to your manuscript, please include the heading and write 'Not applicable' for that section.